\documentclass[epj,referee]{svjour}
\usepackage{latexsym}
\usepackage{epsfig}

\begin{document}
\title{Stochastic Simulations on the Cellular Wave Computers}

\author{M. Ercsey-Ravasz \inst{1,2} \and T. Roska \inst{1}  \and Z. N\'eda \inst{2,3} 
}                     % Do not remove
\institute{P\'azm\'any P\'eter Catholic University, Department of Information Technology,
 HU-1083 Budapest, Hungary \and Babe\c{s}-Bolyai University, Department of Physics, RO-400084 Cluj, Romania \and
Los Alamos National Laboratory, Center for Nonlinear Sciences, NM-87545, Los Alamos, USA}
\date{Received: date / Revised version: date}
\abstract{
The computational paradigm represented by Cellular Neural/nonlinear Networks (CNN) and the CNN Universal Machine (CNN-UM) as a Cellular Wave Computer, gives new perspectives for computational physics. 
Many numerical problems and simulations can be elegantly
addressed on this fully parallelized and analogic architecture. Here we study the possibility of performing
stochastic simulations on this chip. First a realistic random number generator is implemented on the
CNN-UM, and then as an example the two-dimensional Ising model is studied by Monte Carlo type simulations. The results obtained on an experimental version of the CNN-UM with $128 \times 128$ cells are in good agreement with the results 
obtained on digital computers. Computational time measurements suggests that the developing trend of 
the CNN-UM chips - increasing the lattice size and the number of local logic memories - 
will assure an important advantage for the CNN-UM in the near future.   
 % end of PACS codes
} %end of abstract
\maketitle
\section{Introduction}
\label{intro}
Many areas of science and especially physics are prefacing serious problems concerning the computing power of the presently available computers. Solving more and more complex problems, simulating large systems, analyzing huge datasets for which even storing represents a problem, are just a few examples which reminds us that computing power needs to keep up with it's exponential growth, as expressed by Moore's law \cite{moore}. We know however that this process can not continue much further solely with the classical digital computers and new computational paradigms are necessary. Parallel computing, grid computing and quantum computing are just
the most popular examples. The goal of the present article is to make the physicist community aware of a modern and promising trend which is called by computational scientists and engineers Cellular Wave Computers \cite{roska1}. 
This computer is based on the Cellular Neural/nonlinear Network (CNN) and it is experimentally realized by
different physical principles in the architecture of the CNN Universal Machine (CNN-UM). Possibilities 
of performing fast image processing \cite{applications}, solving in an elegant manner partial differential equations 
\cite{pdeI,pdeII} or studying cellular automata models \cite{popdyn,crounse:rand96} on CNN were already studied. Here we argue, that the CNN architecture is also appropriate for Monte Carlo (MC) type simulations on lattice models. As a specific example we study on an experimental version of CNN-UM (the ACE16K chip which has $128 \times 128$ cells) the well-known second-order phase transition in the two-dimensional Ising model. Due to the fact that some simple
operations are not included in this experimental hardware implementation, on this chip the speed of the simulations 
is in the range of modern PC type computers. We will argue however, that the developing trend of this new hardware (2 and 3 layer complex cell CNN-UM architectures, and a powerful new visual microprocessor is coming out at AnaFocus Ltd. soon) could substantially increase the speed of such simulations, assuring an important advantage for CNN computing.

\section{The CNN Universal Machine}
\label{sec:1}

The theory of cellular neural/nonlinear networks (CNN) appeared in 1988 \cite{chua:cnn88}, but the hardware based on this theory, like the CNN Universal Machine (CNN-UM) \cite{roska:cnnum93} are just now developing. 
The CNN-UM is an analogic (analog+logic) computer which has on it's main processor several thousands of interconnected computational units (cells), working parallelly. The CNN-UM can be easily connected to any PC type computer and programmed through a special programming language \cite{homepage}. This new kind of hardware does not  
replace digital computers, but due to it's special structure and architecture it could represent 
an excellent platform for solving some complex problems of physics which demand high computational power.  
CNN-UM is also extremly usefull as a visual or tactile topographic microprocessor.

The standard CNN is composed by $L \times L$ cells placed on a square lattice and interconnected through the $8$ neighbors. Each cell is an electronic circuit in which the most important element is a capacitor. The voltage of this capacitor is called the state value of the cell  $x_{i,j}(t)$. The cell has also an input value (voltage) $u_{i,j}$, which is constant in time and can be defined at the beginning of an operation. The third characteristic of the cell is the output value $y_{i,j}(t)$. This is equivalent with the $x_{i,j}$ state value in a given range. More specifically it is a piece-wise linear function, bounded 
between $-1$ (white) and $1$ (black): $y=f(x)\equiv \frac{1}{2} ( \mid x+1 \mid - \mid x-1 \mid )$

The connections between the cells are realized with voltage-controlled resistors resulting that the state value of each cell depends on the input and output values of the connected neighbors. The state equation of the CNN cells, resulting from the time-evolution of the equivalent circuit (supposing the $8$ neighbor interactions) is the following \cite{chua:cnn88}:
\begin{eqnarray}
\frac{d x_{i,j}(t)}{d t}=-x_{i,j}( t ) + \sum_{k=-1}^{1} \sum_{l=-1}^{1} A_{k,l} y_{i+k,j+l} ( t )+  \label{state} \\
\nonumber + \sum_{k=-1}^{1} \sum_{l=-1}^{1} B_{k,l} u_{i+k,j+l}  + z_{i,j}  
\end{eqnarray}

The coupling between neighbors can be controlled with matrices $A$ and $B$. Within the standard CNN (and on the hardwares realized up to the present days) $A$ and $B$ are the same for all cells. Parameters $z_{i,j}$ are constant values and can vary from cell to cell. The set of parameters $\{A,B,z\}$ is called a template. An operation is performed by giving the initial states of the cells, the input image (the input values of all cells) and by defining a template. The states of all cells will vary parallelly and the result of the operation will be the final steady state of the CNN. Each operation is equivalent with solving a differential equation defined by the template itself, with the extra condition that the state of a cell remains bounded in the $[-1,1]$ region \cite{chuaroskakonyv}.     

The CNN-UM \cite{roska:cnnum93} is a programmable cellular wave computer in which each cell contains additionally a local analog and a logic unit, local analog and logic memories and  a local communication and control unit. Beside these local units, the CNN-UM has also a global analog programming unit which controls the whole system, making it a programmable computer. It can be easily connected to PC type computers and programmed with special languages, for example the analogic macro code (AMC). 

\section{Random number generators on the CNN-UM}

Many applications ideal for the analogic (analog \&logic) architecture of the CNN-UM were already developed and tested. For practical purposes the most promising applications are for image processing, robotics or sensory computing purposes \cite{applications}. 
The CNN architecture seems also promising when considering complex problems in natural sciences. Studies dealing with  partial differential equations (PDE) \cite{pdeI,pdeII,navierstokes,petras} or cellular automata (CA) models \cite{popdyn,crounse:rand96} prove this. Solving partial differential equations is relatively easy and offers the advantage of continuity in time \cite{pdeI}.  Deterministic cellular automata \cite{popdyn} with simple nearest-neighbor rules are also straightforward to implement in the CNN architecture. In physics however, many of the interesting problems deal with stochastic cellular automaton, random initial conditions or other MC methods on lattices (spin problems, population dynamics models, lattice gas models, percolation etc...). 
Developing and proving the efficiency of stochastic simulation techniques on the CNN-UM - using its stored (or algorithmic) programmability - would be thus an important step toward its success.
 
It is known that for a successful stochastic simulation the crucial starting point is a good random number generator (RNG). While computing with digital processors, the "world" is deterministic and discretized, so in principle there is no possibility to generate quickly random events and thus real random numbers. The implemented RNGs are all pseudo-random number generators working with a deterministic algorithm, and it is believed that their statistics approximates well real random numbers. The reproducibility of the pseudo-random numbers can be sometimes an advantage (debugging the code) but in many cases it presents a serious disadvantage. A first advantage of the analog architecture is the possibility to use the the natural noise on the device and to generate real random numbers. 

There are relatively few papers presenting or using RNGs on the CNN-UM \cite{yalcin:rand2004,crounse:rand96,cnnperc}. The known and used ones are all pseudo-random number generators based on chaotic Cellular Automaton (CA) type update rules, generating binary images with $1/2$ probability of the black and white pixels (logical $1$ and $0$, respectively). They were used mainly in cryptography \cite{crounse:rand96} and watermarking on pictures \cite{yalcin:rand2004}. In a recent paper \cite{cnnperc} we presented a realistic RNG by using the natural noise of the chip. An algorithm for generating binary images with any probability of the black pixels was also described. Here we present briefly this realistic RNG and for more details we recommend \cite{cnnperc}. 

The natural noise of the CNN-UM chip is usually highly correlated in space and time, so it can not be used directly to generate random binary values. Our method is based on a chaotic CA perturbed with the natural noise of the chip. The random nature of the noise eliminates the deterministic properties of the chaotic CA.     

As a starting point the relatively simple but efficient chaotic CA, presented by Crounse \textit{et al.} \cite{crounse:rand96} and Yalcin \textit{et al.} \cite{yalcin:rand2004} called the PNP2D was chosen. This chaotic CA is based on the following update rule
\begin{eqnarray}
x_{t+1}(i,j)=(x_t (i+1,j) \vee x_t (i,j+1) ) \oplus  x_t (i-1,j) \oplus \\
\nonumber \oplus x_t (i,j-1) \oplus x_t (i,j), 
\end{eqnarray}
where $i,j$ are the coordinates of the cells, the index $t$ denotes time-steps 
and $x$ is a logic value $0$ or $1$ (representing white and black pixels, respectively). Symbols $\vee$ and $\oplus$ stand for the logical operations or and exclusive-or (XOR), respectively. As described by the authors this chaotic CA is relatively simple and fast, it passed all important RNG tests and shows very small correlations. It generates binary values $0$ and $1$ with the same $1/2$ 
probability independently of the starting condition. It is a good candidate for a pseudo-random number generator and our first goal is to transform it into a realistic RNG. The way to do this is relatively simple. 
After each time step the $P(t)$ result of the chaotic CA is perturbed with a noisy $N(t)$ binary picture (array) so that the final output is given as: $P^{ \prime} (t)=P(t) \oplus N(t)$.
The symbol $\oplus$ stands again for the logical operation XOR, i.e. pixels which are different on the two pictures will become black (logic value $1$). This operation assures that no matter how $N(t)$ looks like, the density of black pixels remains the same $1/2$. Because the used noisy images contain only very few black pixels (logic values $1$) we just slightly sidetrack the chaotic CA from the original deterministic path and all the good properties of the pseudo-random number generator will be preserved. The $N(t)$ noisy picture is obtained by the following simple algorithm. All pixels of a gray-scale image are filled up with a constant value $a$ and a cut is performed at a threshold $a+z$, where $z$ is a relatively small value. In this
manner all pixels which have smaller value than $a+z$ will become white (logic value $0$) and the others black
(logic value $1$). Like all the logic operations this operation can be also easily represented by a CNN template. 
Since the CNN-UM chip is an analog device, there will always be a natural noise on the gray-scale image.  Choosing thus a proper $z$ value one can always generate a random binary picture with few black pixels. These 
$N(t)$ pictures might be strongly correlated and will fluctuate in time. The time-like fluctuations are 
caused by real stochastic processes in the transistor circuits of the chip and can not be thus controlled.  
They are the source of a convenient random perturbation on the chaotic CA, and are responsible for the
realistic nature of the RNG. In case one would need a repeatable series of pseudo-random 
numbers the chaotic CA is simply not perturbed by the $N(t)$ noisy picture.  

Using now $n$ independent random binary images with $1/2$ density of the black pixels, it is possible to 
generate pictures with any $p$ probability of the black pixels ($p$ being a number represented by $n$-bits,
when expressed as a power of $1/2$). For more details see \cite{cnnperc}. 

This RNG and the described algorithms were tested and are properly working on an ACE16K chip which is an experimental version of the CNN-UM with $128 \times 128$ cells.  It is found that the RNG with $p=0.5$ is 
already almost $5$ times faster on the ACE16K than on modern PC type digital computers. Generating images with other $p$ probabilities is of course slower, depending on $n$ (see \cite{cnnperc}). Taking into account thus the natural trend that the lattice size of CNN-UM chips will be growing and that calculations on this chip are totally parallel, these results predict a promising trend. Some codes and movies about the RNGs on the 
ACE16K chip are available on the home-page dedicated to this study \cite{homepage}.  

\section{Studying the Ising model on the CNN-UM} 

Once a properly working RNG is available, Monte Carlo type simulations on two-dimensional lattice-type  
models are possible. Generating random initial conditions for cellular automata models is straightforward and many simple stochastic lattice models can be relatively easily solved \cite{cnnperc}. Here we consider the 
well-known two-dimensional Ising model. Implementing the MC study of this model on the CNN-UM is  
however not trivial. As it will be shown later a straightforward application of the usual Glauber \cite{glauber} or Metropolis \cite{metropolis} algorithms could lead to problems due to the parallel architecture of the computer.     

In the Ising model the spins can have two possible states $\sigma = \pm 1$. On the CNN-UM these states can be mapped on the "black" and "white" states of the cells. Without an external magnetic field the hamiltonian of the system is 
\begin{equation}
H=-J \sum_{<i,j>} \sigma_{i} \sigma_{j},
\end{equation}
$<i,j>$ representing nearest neighbors.
There are many different MC type methods for studying this basic lattice model. 
Most of them like the Metropolis or the Glauber algorithm are of serial nature, meaning that in each step we  update one single spin. Working however parallelly with all spins, could create some
unexpected problems due to the fact that nearest neighbors are updated simultaneously. Imagine for instance an initial state where the spin-values are assigned using a chessboard pattern. This state will have
a zero total magnetization. Let us consider now the zero-temperature case and the Glauber or Metropolis algorithm. Contrary to what is expected, this system will not order in a simple ferromagnetic phase but it will 
continuously switch between the two opposite chessboard patterns. For eliminating the parallel update of the neighbors which causes such problems but still taking advantage of the parallel nature of the computer,
we impose an extra chessboard mask on the system. In each odd (even) step we update parallelly the spins 
corresponding to the black (white) cells of the chessboard mask. For the chosen spins the simple  
Metropolis algorithm is used. It is simple to realize that our method is equivalent with 
the classical serial Metropolis dynamics in which the spins are updated in a well-defined order. Detailed balance and ergodicity is valid, so the obtained statistics should be the right one. 

Implementing the above scheme on the CNN-UM is realized as follows. In each step we first build three 
additional masks: the first marks the spins with $4$ similar neighbors ($\Delta E=8J$), the second one marks the spins with $3$ similar neighbors ($\Delta E=4J$), and the third represents all the other spins for which $\Delta E \leq 0$. Separating these cells is relatively easy using logic operations and some special templates which can shift the images in different directions (for ex. shifting to right can be realized by the template: 
$A=\{0,0,0,0,2,0,0,0,0\}$, $B=\{0,0,0,1,0,0,0,0,0\}$, $z=0$).  We generate two random images with probability $\exp(-8J/kT)$ and $\exp(-4J/kT)$  and we perform an AND operation between the random image and the corresponding mask. After uniting the results of these two and the third mask ($\Delta E \leq 0$) we get a new mask which marks all spins which have to be flipped. Finally we use the chessboard mask and allow only those spins to flip which correspond to black (white) pixels if the time-step is odd (even). The CNN code developed for
studying this problem can be also downloaded from the home-page dedicated to this study \cite{homepage}.
It worth mentioning that cluster algorithms, like the one proposed by Swendsen and Wang \cite{swendsenandwang} or Wolf \cite{wolff},  seem to be also appropriate for the parallel
architecture of the CNN-UM. 

\begin{figure}                                  
\epsfig{file=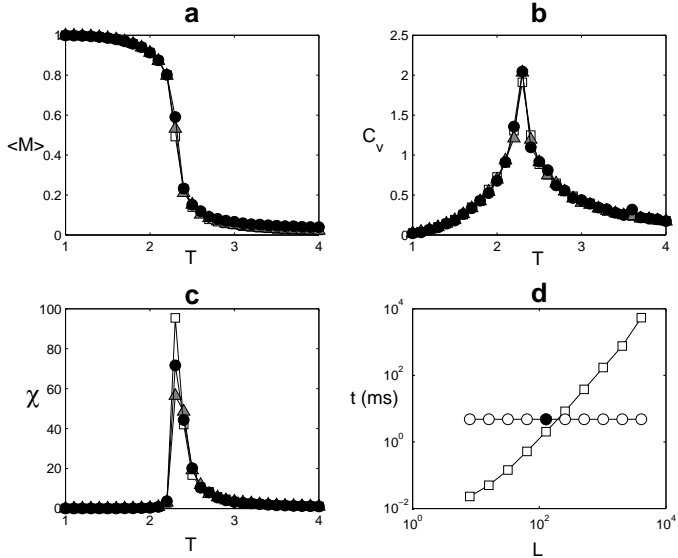,width=0.5\textwidth}
\caption{Average magnetization $M$ (a), specific heat $C_v$ (b) and susceptibility $\chi$ (c) are 
plotted as a function of the temperature $T$ for the classical Metropolis algorithm on a digital computer 
(squares), our parallel algorithm simulated on a digital computer (triangles) 
and the algorithm simulated on the ACE16K CNN-UM chip (circles). Figure (d) compares the simulation 
time  $t$ (in ms) needed for 1 MC step on a Pentium 4 PC with 2.4 GHz (squares) and the CNN-UM (circles) as a function of the lattice size $L$. The filled circle marks the simulation time obtained  on the ACE16K chip ($L=128$). }
\label{fig1}
\end{figure}

Simulation results obtained with the Metropolis type algorithms are sketched on fig. \ref{fig1}. 
On this figure we compare results of (i) the classical Metropolis algorithm on a digital computer, (ii) the results of our parallel algorithm simulated on a digital computer and (iii) the results obtained on the ACE16K chip. 
By plotting the average magnetization, the specific heat and the 
susceptibility as a function of the temperature one can conclude that different results are in good agreement with each other. All simulations were performed on a $128 \times 128$ lattice using free boundary
conditions.

Fig. \ref{fig1}d plots the time needed for 1 MC step as a function of the lattice size $L$. While on a PC type computer this scales as $L^2$, on the CNN-UM the time does not depend on the lattice size (each command is executed in a fully parallel manner on the whole lattice). The time measured on the ACE16K chip with $L=128$ was $4.8 ms$, while on a Pentium 4 PC working on 2.4 GHz under Linux operating system the time needed for $1$ MC step was $2 ms$. For this lattice size the simulations are still faster on the classical digital computers, however considering the trend that the size of the CNN chip (Table 1) will increase in the near future the results are still promising.

{\footnotesize
\begin{center}
\begin{tabular}{|l|c|l|}
\hline
Name & Year & Size \\
\hline
 ---  &   1993  &  $12 \times 12$   \\
\hline
ACE440  & 1995  & $20 \times 22$  \\
\hline
POS48 & 1997 & $48 \times 48$ \\
\hline
ACE4k & 1998 & $64 \times 64$ \\
\hline
CACE1K & 2001 & $32 \times 32 \times 2$ \\
\hline
ACE16K & 2002 & $128 \times 128$ \\
\hline
XENON & 2004 & $128 \times 96 \times 2$ \\
\hline
EYE-RIS & 2005 & $176 \times 144$ \\
\hline
CACE2K & under fabrication & $32 \times 32 \times 3$ \\
\hline

\end{tabular}
\end{center}
\small
TABLE 1. Evolution of the CNN-UM chip, different physical realizations.
From these chips only the ACE16K is commercially available, mass production is expected to begin with the EYE-RIS at the end of 2006.

}

It also worth mentioning here that this ACE16K chip was developed mainly for image processing purposes, the cells have only 2 Local Logic Memories (LLM) and 8 Analog Memories (LAM). While performing logic operations on our binary images we always had to copy the images to the LLMs and  save than the results again to LAMs. These copying processes used around $3/4$ of the processing time. Most of this lost time could be and hopefully will be eliminated in the future by increasing the number of available LLMs. One must also not forget that the CNN-UM was developed mainly for analog signal processing and the main strength of these chips are related to gray scale operators. In that area the proven speed advantage is in about three orders of magnitude \cite{roska1,petras}. 

\section{Conclusions}

In the present study we worked with binary images and we 
exploited mainly the parallel and connectivity features of the CNN.  Our results suggest that the special architecture makes the Cellular Wave Computers very appropriate for simulating lattice models and it's natural noise can be effectively used in stochastic simulations. The ongoing developing process of this hardware is expected to increase the number of cells and local memories, and also  three-dimensional chips with more layers of cells are expected to appear. This would assure an important advantage for these chips in the near future. We
think that CNN computing could be effectively used in computational physics for supplementing digital computers in some complex and time-consuming problems.

%$%%$$$$$$$$$$$$$$$$$$$$ Hogy kell beirni hogy ne legyen szama ennek a section-nek???
\section{Acknowledgments}
The support of the Jedlik Laboratories of the P. P\'azm\'any Catholic 
University is gratefully acknowledged.

\end{document}